\documentclass[sigconf]{acmart}

\AtBeginDocument{%
  }

\setcopyright{none}
\acmConference[arXiv Preprint]{}{April 2026}

\usepackage{multirow}
\usepackage{tcolorbox}

\begin{document}

\title{VisionClaw: Always-On AI Agents Through Smart Glasses}

\renewcommand{\authorsaddresses}{}
\author{Xiaoan Liu}
\authornote{Equal contribution.}
\affiliation{%
  \institution{University of Colorado Boulder}
  \city{Boulder}
  \state{Colorado}
  \country{USA}}
\email{xiaoan.liu@colorado.edu}

\author{DaeHo Lee}
\authornotemark[1]
\authornote{This work was done during a visiting scholarship at CU Boulder.}
\affiliation{%
  \institution{Gwangju Institute of Science and Technology}
  \city{Gwangju}
  \country{Republic of Korea}}
\email{leedaeho@gm.gist.ac.kr}

\author{Eric J. Gonzalez}
\affiliation{%
  \institution{Google}
  \city{Seattle}
  \state{Washington}
  \country{USA}}
\email{ejgonz@google.com}

\author{Mar Gonzalez-Franco}
\affiliation{%
  \institution{Google}
  \city{Seattle}
  \state{Washington}
  \country{USA}}
\email{margon@google.com}

\author{Ryo Suzuki}
\affiliation{%
  \institution{University of Colorado Boulder}
  \city{Boulder}
  \state{Colorado}
  \country{USA}}
\email{ryo.suzuki@colorado.edu}

\renewcommand{\shortauthors}{Liu et al.}
\newcommand{\rev}[1]{{\color{blue}{#1}}}
\newcommand{\todo}[1]{{\color{red}{#1}}}

\begin{abstract}
We present VisionClaw, an always-on wearable AI agent that integrates live egocentric perception with general-purpose agentic task execution. Running on Meta Ray-Ban smart glasses, VisionClaw continuously perceives real-world context and enables in-situ, speech-driven action delegation via Gemini Live and OpenClaw AI agents. This allows users to directly execute tasks through the smart glasses, such as adding real-world objects to an Amazon cart, generating notes from physical documents, receiving meeting briefings on the go, creating calendar events from posters, or controlling IoT devices. We evaluate VisionClaw through a controlled laboratory study (N=12). Results show that our system enables 13--37\% faster task completion and 7--46\% lower perceived difficulty compared to an agent-only and always-on-only baselines, respectively. Beyond performance gains, an autobiographical deployment study (N=4) over an average of 13.8 active days per user, with a total of 555 interactions over 25.8 hours, reveals six usage categories and four emergent interaction patterns, which suggest a new paradigm of wearable AI agents for ubiquitous, situated, and ambient human-AI interaction.
\end{abstract}



\begin{teaserfigure}
  \includegraphics[width=\textwidth]{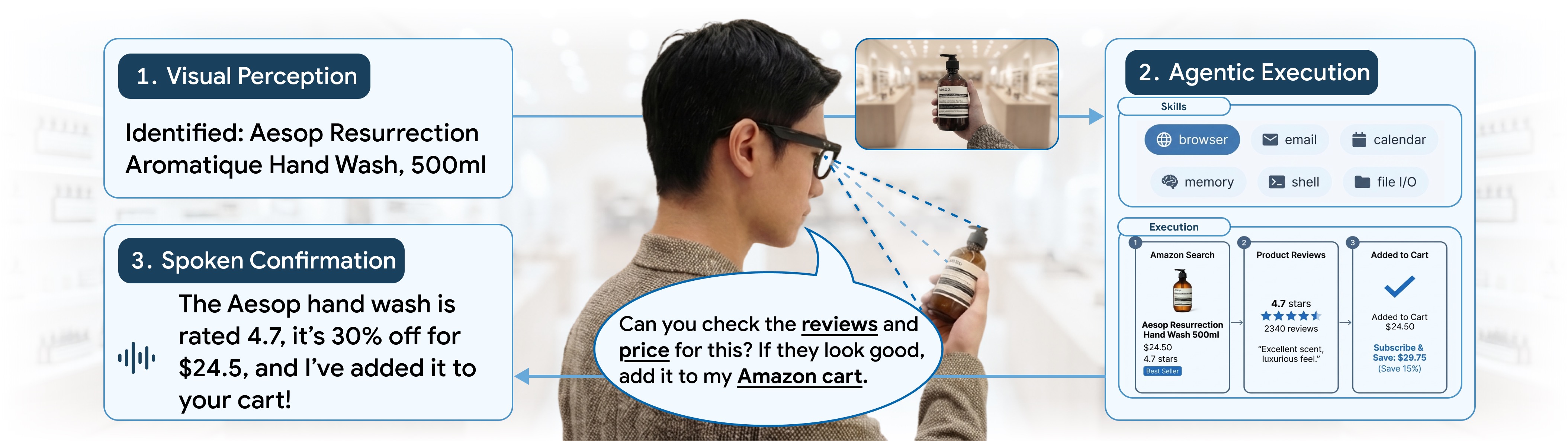}
  \caption{VisionClaw integrates always-on egocentric perception with agentic task execution on smart glasses. A user holding a product says ``Can you check the reviews and price for this? If they look good, add it to my Amazon cart.'' The system 1) identifies the product through visual perception, 2) autonomously executes a multi-step browser workflow to find and add the item on Amazon, and 3) confirms completion through spoken feedback---all without the user touching a screen.}
  \Description{A person wearing smart glasses in a retail store holds an Aesop hand wash bottle. Three overlaid panels show the system workflow: visual perception identifying the product, agentic execution adding it to an Amazon cart, and spoken confirmation of the completed action.}
  \label{fig:teaser}
\end{teaserfigure}

\maketitle

\section{Introduction}

The vision of an always-on AI agent that accompanies users throughout their daily lives has long appeared in both science fiction and technology forecasting, from the ambient AI companion in the movie \textit{Her}~\cite{her2013} to Apple's \textit{Knowledge Navigator}~\cite{appleknowledgenavigator1987}. Recent advances in large language models and multimodal AI systems~\cite{openai2023gpt4v, geminilive2024} have brought this vision closer to reality. However, realizing a general-purpose AI assistant that is both \textit{always-on} and \textit{capable of acting} in the user's real-world context is still challenging.

Two parallel research directions have emerged toward this goal. On one hand, recent work on autonomous AI agents has demonstrated increasing capabilities in executing digital tasks such as web navigation, email composition, memory retrieval, and calendar management~\cite{zhang2025appagent, wang2024mobile, wu2024copilot, zhou2023webarena, openclaw2025}. Yet these systems remain primarily designed for desktop or smartphone environments, and thus rely heavily on \textit{screen-based interaction}~\cite{vu2024gptvoicetasker}. On the other hand, AI-powered smart glasses can see and hear the user's surroundings~\cite{konrad2024gazegpt, cho2025persistent}, but these systems are largely limited to simple \textit{question answering}, thus task execution capability is highly limited~\cite{huang2025vinci, lee2024gazepointar, cai2025aiget, zulfikar2024memoro}. As a result, neither approach alone delivers the experience of \textit{general-purpose always-on AI agents} that can both perceive the physical world and take action. Moreover, most of the prior work~\cite{lee2025sensible, pu2025promemassist, yang2025proagent, fang2025mirai} has not explored a longitudinal deployment study to investigate how such agentic wearable systems are actually used in everyday settings.

In this paper, we introduce VisionClaw, an open-source always-on wearable AI agent system that bridges this gap by connecting real-time egocentric perception from smart glasses with executable AI agents. VisionClaw enables users to initiate tasks directly from what they are currently seeing through voice interactions by leveraging Meta Ray-Ban glasses connected to Gemini Live~\cite{geminilive2024} for multimodal perception and OpenClaw~\cite{openclaw2025} as an agentic backend. For example, users can add a product they are looking at to an online shopping cart, generate an email based on a physical document, receive meeting briefings on the go, create a calendar schedule based on an event poster, or control connected IoT devices directly through glasses. By combining always-on wearable perception with agentic task execution, VisionClaw reveals new use cases and interaction patterns in which everyday tasks emerge opportunistically from the user's visual context.

To investigate how such always-on AI agents affect usability and real-world interaction, we conducted two complementary studies. First, we performed a controlled user study (N=12) comparing VisionClaw with two conditions: 1) always-on only (Meta Ray-Ban and Gemini Live) and 2) agent only (OpenClaw with a smartphone) across four tasks requiring reference to objects and documents in the physical environment. Second, we conducted a longitudinal autobiographical deployment study with four authors who used VisionClaw in their daily lives. Across the field deployment, participants engaged in 555 voice-initiated interactions over 55 active participant-days throughout 25.8 total hours (5--19 days per participant with an average of 13.8 days, averaging 10.1 interactions per day), revealing how an always-on wearable agent integrates into everyday routines.

Our findings from both studies highlight key results. The user study shows that VisionClaw enables 13--37\% faster task completion and 7--46\% lower perceived difficulty compared to agent-only and always-on-only baselines, respectively, with significantly lower cognitive workload (NASA-TLX mental demand, temporal demand, and frustration, $p$<0.05). The deployment study reveals six use case categories---\textit{Communicate} (14\%), \textit{Retrieve} (30\%), \textit{Save} (16\%), \textit{Recall} (12\%), \textit{Shop} (19\%), and \textit{Control} (9\%)---where general-purpose agents enable open-ended, multi-turn conversations that chain across categories within a single session. Together, these findings suggest that always-on agentic systems reshape interaction from discrete, command-based usage to continuous, context-driven engagement, where perception, memory, and action are tightly integrated and interactions emerge opportunistically over time. Based on these findings, we discuss implications for always-on wearable AI agents: privacy risks of continuous capture coupled with autonomous action, memory architectures for unbounded life streams, and the challenge and opportunity of designing agents that recede into calm, ambient assistance rather than demanding foreground attention.

Finally, our contributions are as follows:

\begin{itemize}
\item The architecture and open-source implementation of VisionClaw, an always-on wearable AI agent that integrates real-time perception with general-purpose task execution.
\item A controlled laboratory study (N=12) evaluating our system, compared to always-on only and agent only conditions.
\item An autobiographical deployment study (N=4) revealing emergent use cases, interaction patterns, and properties for always-on wearable AI agents.
\end{itemize}

\section{Related Work}
\subsection{LLM-Based Agents for Task Execution}

Large language models can now invoke external tools to carry out multi-step digital tasks on behalf of users. Foundational work established autonomous tool invocation~\cite{schick2023toolformer} and reasoning-and-acting loops~\cite{yao2022react} as common agent design patterns. These have since been extended to multi-agent collaboration~\cite{hong2023metagpt} and to increasingly broad action spaces, from web navigation~\cite{zhou2023webarena} and desktop applications~\cite{wu2024copilot} to unified service layers spanning email, browsing, and calendar management~\cite{openclaw2025}. Within HCI, a growing body of work has explored how such agents operate in everyday interaction environments. One direction has focused on mobile and GUI automation, where agents learn to operate smartphone applications through human-like gestures 
\cite{zhang2025appagent, wang2024mobile},  high-resolution visual encoding \cite{hong2024cogagent}, or voice-initiated task flows \cite{vu2024gptvoicetasker}. A second direction has addressed user agency, where agents pause at decision points to solicit user preferences before proceeding \cite{peng2025morae}.

While these systems demonstrate strong capabilities for digital task execution, they are primarily designed around \textit{screen-based interactions} and lack direct awareness of the user's physical surroundings. Recent work has highlighted the barriers scientists and professionals face when attempting to use AI for embodied physical tasks that go beyond the desk~\cite{hou2026beyond}, underscoring the limited support current agents offer for \textit{situated} interaction where tasks are initiated and grounded in the user's real-world context.

\subsection{Wearable AI and Smart Glasses}

Wearable computing has a long history of exploring always-on, hands-free systems for augmenting human capabilities, from proactive context-aware information retrieval~\cite{rhodes1997wearable} and passive wearable cameras as memory aids~\cite{hodges2006sensecam} to cloudlet-based real-time task guidance through head-worn displays~\cite{ha2014towards}. Smart glasses and AR headsets have recently become a primary platform for AI-powered wearable interaction. One active area is multimodal input for hands-free interaction, where systems combine eye gaze~\cite{konrad2024gazegpt, rekimoto2025gazellm, zhang2026gazeify}, gestures~\cite{lee2024gazepointar, hu2025gesprompt}, and voice~\cite{kim2026speech} to disambiguate references and ground utterances in the physical world. Researchers have also explored situated task guidance, generating step-by-step AR instructions grounded in the user's environment \cite{liu2023instrumentar, huh2025vid2coach, nguyen2025install, zhang2025following, zhao2025guided}, with some leveraging wearable sensing for proactive guidance~\cite{mahfuz2026proactive, arakawa2024prism}. Additional work has explored contextual augmentation, including live visual descriptions for accessibility~\cite{chang2024worldscribe, lee2024cookar}, spatial memory anchoring~\cite{kim2026speechless}, augmented object intelligence~\cite{dogan2024augmented}, speech-driven conversational assistant~\cite{jadon2024augmented, olwal2020wearable}, and everyday knowledge discovery on smart glasses~\cite{cho2025persistent, cai2025aiget}.

More recently, several smart glasses have introduced agent-like capabilities, moving beyond passive assistance toward autonomous context-sensitive behavior---autonomously deciding what to assist with based on egocentric sensing~\cite{lee2025sensible, xu2023xair}, leveraging visual data for task understanding~\cite{li2025satori, wang2023holoassist}, using multi-sensor context for proactive LLM-powered assistance~\cite{yang2025proagent, fang2025mirai}, and surfacing memories or timing interventions based on the context and cognitive state~\cite{pu2025promemassist, kim2026memento, zulfikar2024memoro}. Survey and workshop efforts~\cite{tang2025llm, suzuki2023xr, suzuki2025everyday} have further mapped the research agenda for AI-enabled wearable and AR systems.

However, these agent-like wearable systems remain \textit{specialized} for particular tasks such as question-answering, memory and information retrieval, and step-by-step guidance. None provides a \textit{general-purpose} task execution capability, nor have longitudinal studies examined how such general-purpose always-on agents are used in everyday settings.

\subsection{Context-Aware Egocentric Perception}

Advances in vision-language models have extended egocentric perception, building on context-aware computing~\cite{abowd1999towards, dey2001understanding}. Large-scale egocentric video benchmarks~\cite{grauman2022ego4d, grauman2024ego}, sensing platforms~\cite{engel2023project}, and video-language pretraining~\cite{lin2022egocentric} have together enabled systems for long-context question answering~\cite{yang2025egolife, goletto2024amego}, real-time vision-language assistance~\cite{huang2025vinci}, personalized streaming understanding~\cite{zheng2026pearl}, and attention-aware content adaptation~\cite{pei2025attentionar}. Benchmarks have also begun coupling perception with digital task completion~\cite{li2024omniactions, yu2026ego2webwebagentbenchmark, zhao2025less}. Meanwhile, always-on egocentric sensing raises privacy concerns~\cite{hoyle2014privacy, denning2014situ, bhardwaj2024focus}, motivating privacy-control mechanisms~\cite{zhang2026visguardian, li2025egoprivacy, wang2026mind}. These advances have expanded what egocentric systems can perceive and recognize, but the interaction design space---what tasks naturally arise in everyday settings and how interaction patterns change over time---remains largely unexamined from an HCI perspective. Our longitudinal deployment study provides empirical grounding for these open questions.

\section{System Design}

VisionClaw was designed around three goals:

\begin{enumerate}
  \item \textbf{Persistent environmental perception.} The system should continuously share the user's visual and auditory context, rather than requiring explicit context-providing actions such as taking a photo or typing a description.
  \item \textbf{Real-time natural interaction.} The user should be able to converse with the system through natural speech and gestures, receiving spoken responses with low perceived latency. The interaction should feel conversational rather than command-based.
  \item \textbf{Agentic execution.} When the user expresses an actionable intent, the system should be able to carry it out by invoking real-world services, without requiring the user to switch to a phone or computer.
\end{enumerate}

\subsection{Architecture Overview}

\begin{figure*}[t]
  \centering
  \includegraphics[width=\textwidth]{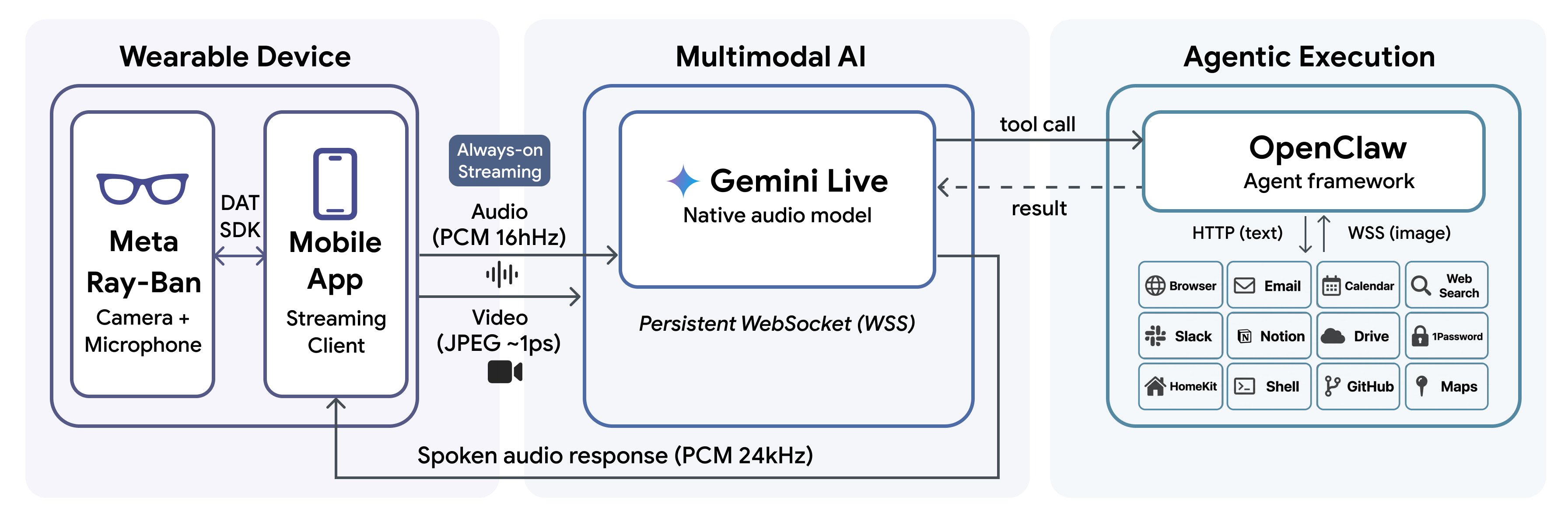}
  \caption{System architecture of VisionClaw. The wearable device layer captures audio and video from Meta Ray-Ban smart glasses via the DAT SDK and streams them through an phone app. These always-on streams are sent to the Gemini Live API over a persistent WebSocket connection. Gemini processes the multimodal input and either responds with spoken audio or issues tool calls routed to OpenClaw for execution via dual HTTP and WebSocket channels.}
  \Description{A system architecture diagram with three color-coded containers arranged left to right. The green container labeled Wearable Device contains Meta Ray-Ban and iPhone App boxes side by side, connected by DAT SDK. The blue container labeled Multimodal AI contains a Gemini Live box with capture\_photo and execute tool tags. The red container labeled Agentic Execution contains an OpenClaw box and six category boxes: Retrieve, Shop, Communicate, Control, Save, and Recall. Horizontal arrows show audio and video streams flowing from the wearable device to Gemini, tool calls and results between Gemini and OpenClaw, and a horizontal blue arrow at the bottom showing spoken audio response returning to the user.}
  \label{fig:pipeline}
\end{figure*}

\autoref{fig:pipeline} illustrates the system architecture. VisionClaw consists of three layers: 1) a sensory input layer that captures audio and video from Meta Ray-Ban smart glasses; 2) a multimodal AI layer powered by the Gemini Live API that processes the streaming sensor data and generates responses; and 3) an agentic execution layer powered by OpenClaw that carries out real-world tasks when the AI model issues tool calls.

\subsection{Sensory Input Layer}

The sensory input layer captures the user's perceptual context through two channels: audio and video. In glasses mode, audio is captured from the Meta Ray-Ban built-in microphone, and video frames are received from the glasses camera at 24 frames per second. Video frames are throttled to approximately 1 frame per second and compressed to JPEG at 50\% quality before transmission, balancing visual fidelity against bandwidth and API cost. The user can also enable an audio-only mode to reduce battery drain and mitigate unstable network connections during always-on video capture for longer outdoor usage.

\subsection{Multimodal AI Layer}

VisionClaw leverages the Gemini Live API, accessed through a persistent WebSocket connection. The system uses the gemini-2.5-flash-native-audio-preview model, which accepts interleaved audio chunks and JPEG frames and produces spoken responses. A key property of this approach is that audio understanding is native: the model processes raw audio directly rather than first transcribing it to text. This preserves prosodic cues and enables lower-latency interaction compared to a pipeline of separate speech-to-text, language model, and text-to-speech components. The system prompt instructs Gemini to behave as a context-aware assistant that can see through the user's glasses and hear their surroundings. When the model determines that a user request requires executing a real-world action, it generates a function call (e.g., \texttt{execute(task: "add eggs to my shopping list")}), which is intercepted by the client and forwarded to the agentic execution layer.

\subsection{Agentic Execution Layer}

When the Gemini model issues a tool call, the VisionClaw client routes it to OpenClaw~\cite{openclaw2025}, an open-source agentic framework running as a local gateway on a Mac or a virtual private server (VPS). OpenClaw provides various skills, including email composition, web browsing, memory retrieval, calendar management, messaging, and file operations. Each skill is implemented as an autonomous sub-agent capable of multi-step task execution. For example, the ``browser automation'' skill can navigate to Amazon, search for a product, and add it to the user's cart through browser automation. The execution result is returned to the VisionClaw client, which sends it back to Gemini as a tool response. Gemini then generates a spoken response for the user. This completes the see-and-act loop: the user perceives real-world objects, speaks a request, and receives spoken confirmation that the action has been carried out.

\section{Implementation}

VisionClaw is implemented as native mobile applications for both iOS and Android. The iOS version is built in Swift using SwiftUI, targeting iOS 17.0 and above, and has been tested on iPhone 15 Pro and iPhone 16 Pro. The Android version is built in Kotlin, and has been tested on Google Pixel and Samsung Galaxy. Both versions pair with Meta Ray-Ban smart glasses. The application is open-source and publicly available at \url{https://github.com/Intent-Lab/VisionClaw}.

\subsection{Glasses Communication}

Communication with the Meta Ray-Ban smart glasses is managed through Meta's Device Access Toolkit (DAT) SDK~\cite{metadat2024}. The SDK requires that developer mode be enabled on the glasses through the Meta AI companion app. Once paired, the DAT SDK provides a \texttt{videoFramePublisher} that delivers camera frames at 24~fps, as well as audio streaming capabilities. The \texttt{WearablesViewModel} manages the device lifecycle, including registration, pairing, and streaming state.

\subsection{Gemini WebSocket Client}

The \texttt{GeminiLiveService} manages the WebSocket connection to the Gemini Live API. Upon session initiation, the client sends a setup message containing the model identifier, system prompt, generation parameters, and tool declarations. Audio chunks and JPEG frames are then sent as interleaved messages. Incoming messages are parsed to distinguish between audio response chunks, text transcriptions, and tool call requests. The \texttt{GeminiSessionViewModel} orchestrates the session lifecycle, including connection management, transcript accumulation, and tool call wiring.

\subsection{Audio Pipeline}

The \texttt{AudioManager} handles bidirectional audio. For input, it configures an \texttt{AVAudioEngine} tap that captures microphone audio, resamples it to 16~kHz mono PCM Int16, and buffers it into 100~ms chunks for transmission. For output, it maintains a playback queue that receives PCM 24~kHz audio chunks from Gemini and plays them sequentially through the device speaker. The audio session category and mode are configured based on whether the system is operating in glasses mode or phone mode.

\subsection{Tool Call Routing}

The \texttt{ToolCallRouter} receives function call messages from Gemini and dispatches them to the OpenClaw gateway via the \texttt{OpenClawBridge}. The bridge sends HTTP POST requests to the gateway (default endpoint: \texttt{http://<local-ip>:18789}) with bearer token authentication. It tracks in-flight tasks and handles timeouts. Results are formatted as tool response messages and sent back to Gemini over the WebSocket connection.

\section{User Study}
\label{sec:evaluation}

\subsection{Method}

\subsubsection*{\textbf{Study Design}}

We compared three interaction conditions in a within-subjects design, 
while keeping the primary interaction modality (real-time, voice-based interaction) consistent across conditions and isolating the effects of perception and agentic execution.

\noindent
\textbf{1) Always-On Only (Meta Ray-Ban and Gemini Live):} Participants used glasses running Gemini Live. They could use the camera and voice interaction on the glasses to ask questions about the visual context. However, the system could not execute tasks autonomously, thus participants had to perform execution manually by switching to phones (e.g., adding items to a shopping cart or writing notes with transcribed chat logs). We provide the full system prompt in Appendix~\ref{sec:always-on-prompt}.

\noindent
\textbf{2) Agent Only (Smartphone with OpenClaw):} Participants used a smartphone-based OpenClaw chat interface capable of executing tasks (e.g., web browsing, email composition, note creation). The system lacked visual perception, so participants had to explicitly provide context via text or voice (speech-to-text transcription).

\noindent
\textbf{3) Always-On and Agent (VisionClaw):} Participants wore Meta Ray-Ban smart glasses running VisionClaw. The system continuously streamed egocentric camera input and supported voice interaction. When users issued actionable requests, the system executed tasks through OpenClaw's agent tools. We provide the full system prompt in Appendix~\ref{sec:visionclaw-prompt}

\begin{table}[h]
\centering
\small
\begin{tabular}{lcc}
\toprule
Condition & Visual Perception & Agentic Execution \\
\midrule
Always-On Only & \checkmark & \\
Agent Only &  & \checkmark \\
Always-On and Agent & \checkmark & \checkmark \\
\bottomrule
\end{tabular}
\caption{Capabilities supported in each condition.}
\Description{A table comparing three experimental conditions—Always-On Only, Agent Only, and Always-On and Agent—across two capabilities: Visual Perception and Agentic Execution. Always-On Only includes Visual Perception but not Agentic Execution, Agent Only includes Agentic Execution but not Visual Perception, and Always-On and Agent includes both capabilities.}
\label{tab:conditions}
\end{table}

\subsubsection*{\textbf{Participants}}

We recruited 12 participants (9 male, 3 female; age: M = 27.8, SD = 6.16) from a local community. Participants reported high familiarity with AI assistants (M = 6.0/7), but moderate familiarity with AI Agents (M = 3.92/7) and voice assistants (M = 3.92/7). One participant had prior experience with AI smart glasses. Each session lasted approximately 90 minutes, and each was compensated \$30.

\subsubsection*{\textbf{Tasks}}

\begin{figure}[h]
  \centering
  \includegraphics[width=\linewidth]{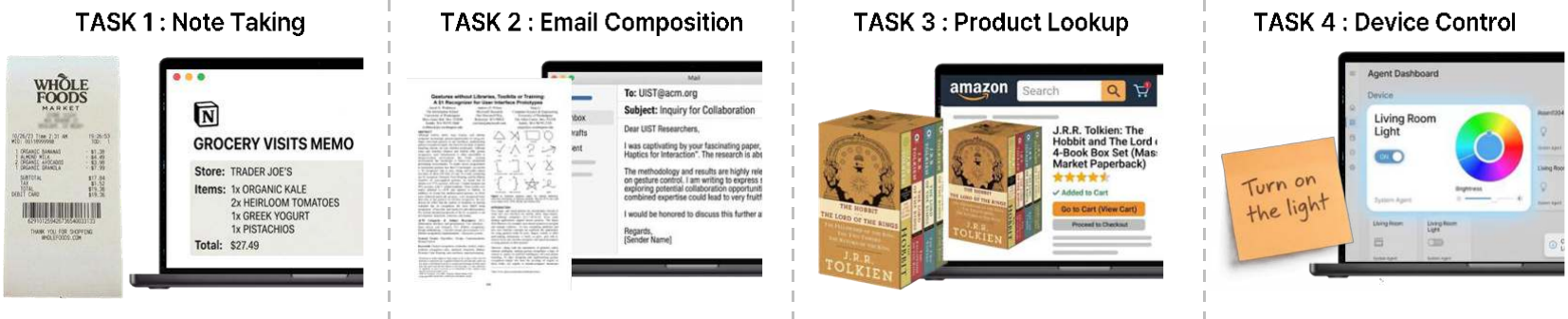}
  \caption{Overview of the four tasks used in the study}
  \Description{A sequence of example tasks used in the study. From left to right: (1) a physical receipt is converted into a structured grocery memo on a laptop (note taking), (2) a printed academic paper is used to compose an email draft (email composition), (3) a book is identified and compared on an online shopping interface (product lookup), and (4) a handwritten note instructing “Turn on the light” corresponds to a smart home control interface (device control). The figure illustrates how tasks are grounded in real-world physical artifacts and mapped to digital actions.}
  \label{fig:task_desc}
\end{figure}

We designed four tasks (\autoref{fig:task_desc}) grounded in physical artifacts to simulate everyday situations in which users observe real-world objects and initiate digital actions.

\noindent
\textbf{1) Note Taking:} Given a printed receipt, create a note in Notion containing the store name, items purchased, and total price.

\noindent
\textbf{2) Email Composition:} Given a printed academic paper, compose a draft email in Gmail requesting research collaboration with the paper’s first author, including a proposed collaboration intent.

\noindent
\textbf{3) Product Lookup:} Given a printed book cover, retrieve the Amazon review score. If the score exceeds 4.5, add the item to the shopping cart.

\noindent
\textbf{4) Device Control:} Control a smart light bulb by turning it on, off, or changing its color based on a given instruction.

\subsubsection*{\textbf{Procedure}}

Participants completed all four tasks under each of the three conditions. The order of conditions was counterbalanced using a Latin square to reduce order effects. To familiarize participants with the systems, each condition began with a brief tutorial followed by a practice task. In the practice session, participants used all three systems to complete a calendar scheduling task based on a printed event poster, ensuring basic understanding of interaction flow and system capabilities.

To mitigate learning effects across repeated tasks, we prepared three distinct instances of each task (e.g., different receipts, flyers, and book covers) and assigned them across conditions in a counterbalanced manner. This ensured that participants did not encounter the same task content more than once.

\subsubsection*{\textbf{Measurement}}

For each task, we recorded task completion time, task success, and perceived task difficulty. After completing the tasks for each condition, participants filled out questionnaires. We collected subjective workload using the NASA-TLX (0–20 scale), along with six self-authored 7-point Likert items: \textit{1) perceived control}, \textit{2) reliability}, \textit{3) trust}, \textit{4) ease of use}, \textit{5) usefulness}, and \textit{6) confidence} (see Appendix~\ref{sec:self-authored} for details). After all three conditions, participants completed a short semi-structured interview.
\subsection{Results}

\subsubsection*{\textbf{Task Completion Time}}

\begin{figure}[h]
  \centering
  \includegraphics[width=\linewidth]{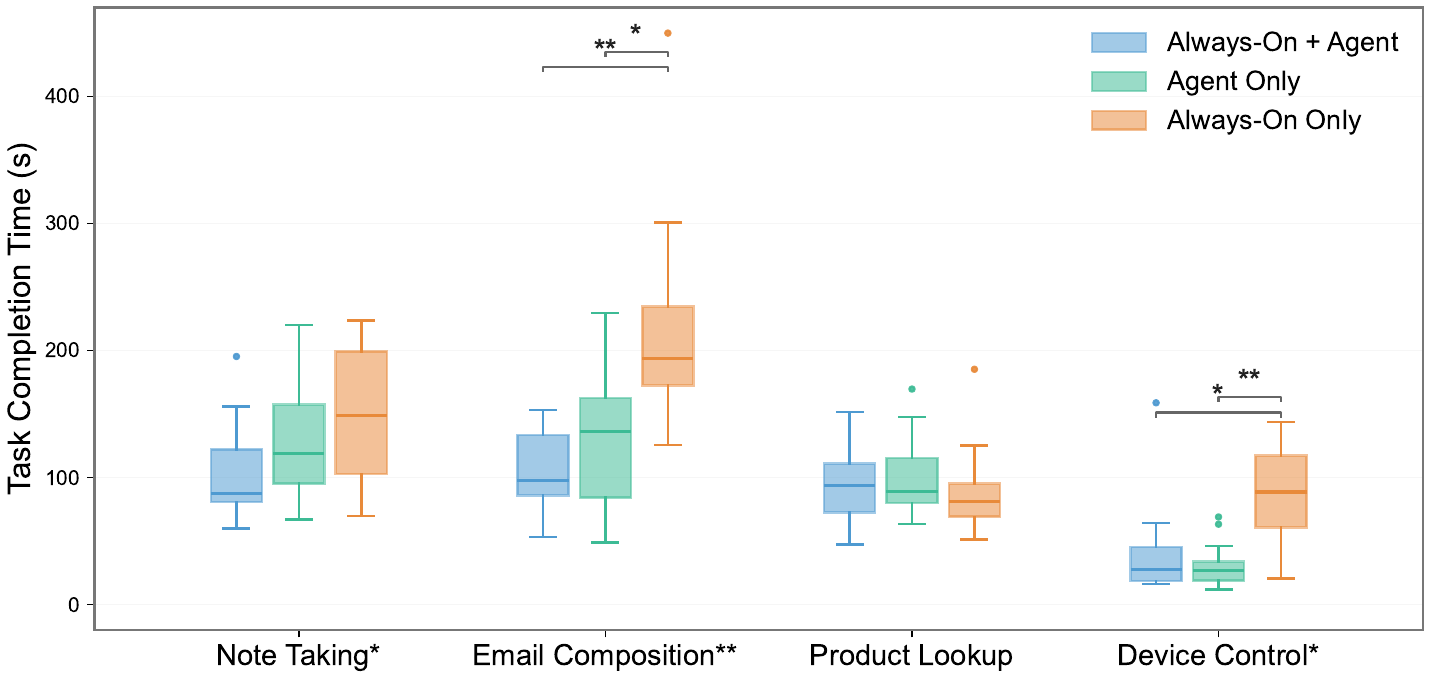}
  \caption{Task completion time. Asterisks next to labels indicate significance from Friedman tests, and bracketed asterisks indicate significance from Wilcoxon signed-rank tests. (* p < .05, ** p < .01.)}
  \Description{Box plots showing task completion time (in seconds) across four tasks—Note Taking, Email Composition, Product Lookup, and Device Control—for three conditions: Always-On + Agent, Agent Only, and Always-On Only. Boxes indicate median and interquartile range, with individual data points overlaid. Significant differences are observed in Note Taking, Email Composition, and Device Control, where Always-On + Agent generally achieves faster completion times compared to the other conditions, particularly outperforming Always-On Only.}
  \label{fig:task_completion}
\end{figure}

\autoref{fig:task_completion} shows the task completion time across all four tasks. A Friedman test revealed a significant effect of condition on task completion time for \textit{Note Taking} ($\chi^2(2)=7.17, p=.028$), \textit{Email Composition} ($\chi^2(2)=10.17, p=.006$), and \textit{Device Control} ($\chi^2(2)=8.00, p=.018$), but not for \textit{Product Lookup} ($\chi^2(2)=1.50, p=.472$). Post-hoc comparisons showed that \textit{VisionClaw} was significantly faster than the \textit{Always-On Only} condition in \textit{Email Composition} ($p=.0029$), and that both \textit{VisionClaw} ($p=.0322$) and \textit{Agent Only} ($p=.0073$) were significantly faster than the \textit{Always-On Only} condition in \textit{Device Control}.

Participants reported that tasks involving substantial text input or transcription, such as \textit{Note Taking} and \textit{Email Composition}, were more demanding, which contributed to longer completion times. In the \textit{Always-On Only} condition, participants highlighted that they \textit{``have to do things manually in the end''} (P1), increasing the overall effort and time required. Additionally, several participants pointed out increased interaction overhead, noting that they \textit{``still have to say everything''} (P12) and \textit{``need to give all the information by talking to it''} (P11) in \textit{Agent Only} condition.

In contrast, \textit{Product Lookup} involved less text and more familiar interactions, resulting in relatively consistent time across conditions. For \textit{Device Control}, some participants required additional time to learn how to execute the task manually, particularly in the \textit{Always-On Only} condition (P2, P8).

\subsubsection*{\textbf{Task Success Rate}}

Task success rates were high across systems (\textit{VisionClaw}: $85.4\%$, \textit{Agent Only}: $89.6\%$, \textit{Always-On Only}: $97.9\%$ at participant-level means), and we found no significant between-system differences in success rate (Friedman: $\chi^2(2)=2.63$, $p=.269$; all pairwise Wilcoxon tests did not show significance). 

Task success was determined based on whether the essential task requirements were correctly fulfilled. Minor typographical errors were tolerated and counted as successful outcomes. However, responses were marked as failures if critical information was missing, incorrect, or if errors were substantial enough to make the result difficult to understand or use. Notably, the note-taking task showed a relatively lower success rate with VisionClaw (58\%), primarily due to challenges in capturing small or visually constrained artifacts such as receipts, which often led to recognition errors.

\begin{figure}[h]
  \centering
  \includegraphics[width=\linewidth]{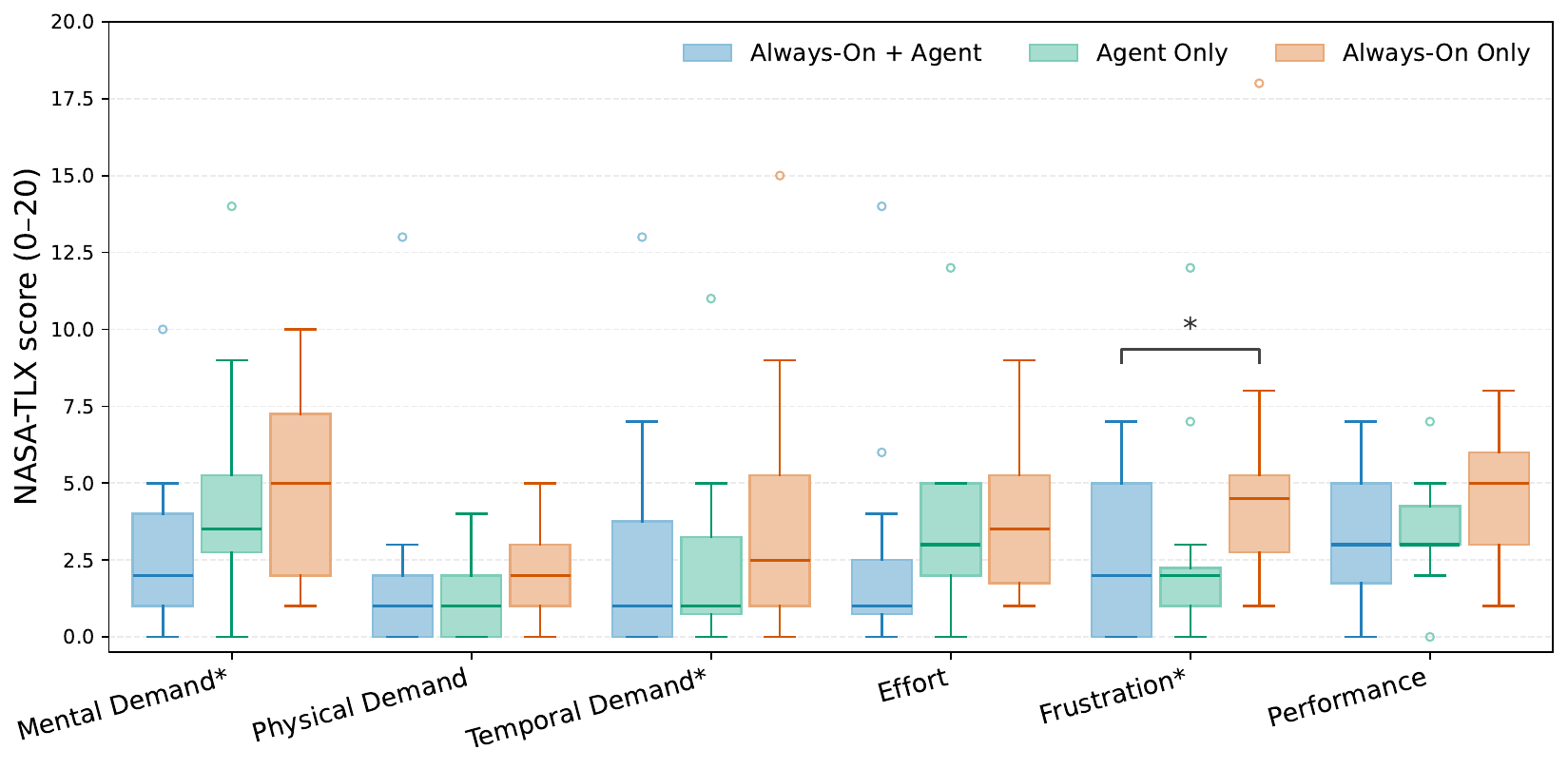}
  \caption{NASA-TLX. Asterisks next to labels indicate significance from Friedman tests, and bracketed asterisks indicate significance from Wilcoxon signed-rank tests. Lower scores indicate better performance. (* p < .05, ** p < .01)}
  \Description{Box plots of NASA-TLX workload scores (0–20 scale) across six dimensions: Mental Demand, Physical Demand, Temporal Demand, Effort, Frustration, and Performance, for three conditions (Always-On + Agent, Agent Only, Always-On Only). Lower scores indicate lower workload (except Performance, where lower indicates better perceived performance). Significant reductions in Mental Demand, Temporal Demand, and Frustration are observed for Always-On + Agent compared to the other conditions, indicating reduced cognitive load.}
  \label{fig:nasa_tlx}
\end{figure}

\subsubsection*{\textbf{Subjective Ratings}}

Subjective ratings revealed significant effects of condition on workload and user experience measures. For NASA-TLX, we observed significant effects on \textit{mental demand} ($\chi^2(2)=9.80, p=.007$), \textit{temporal demand} ($\chi^2(2)=6.41, p=.040$), and \textit{frustration} ($\chi^2(2)=6.00, p=.049$), with \textit{VisionClaw} showing lower scores than both \textit{Always-On Only} and \textit{Agent Only}. Other dimensions were not significant. Participants attributed frustration to occasional perception errors, noting that the system could \textit{``get information wrong''} (P2) or \textit{``fail to capture details accurately''} (P4). Perceived difficulty was generally lower for \textit{VisionClaw} than \textit{Always-On Only}, suggesting reduced interaction overhead. 

\begin{figure}[h]
  \centering
  \includegraphics[width=\linewidth]{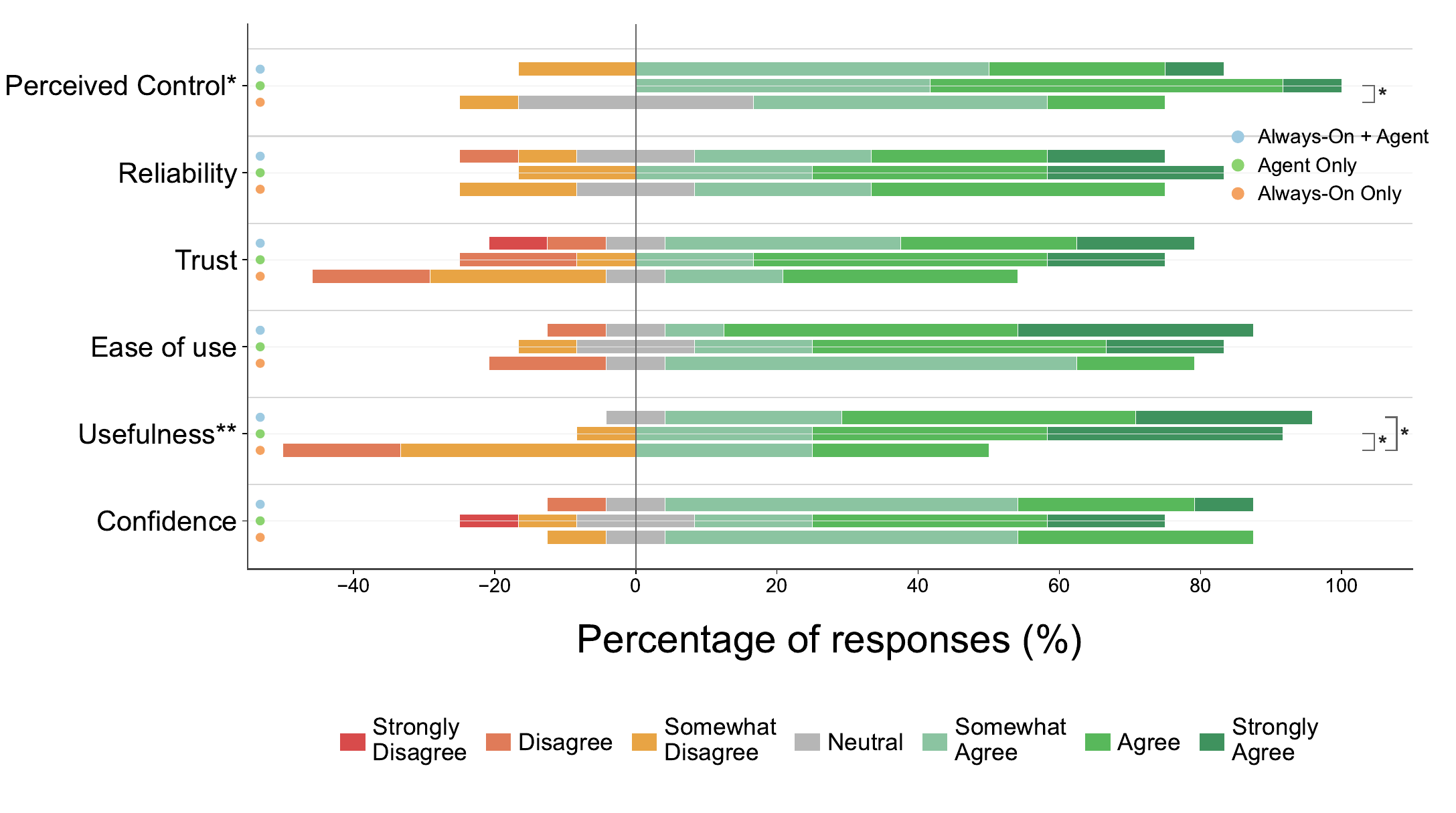}
  \caption{Self-authored questionnaire. Asterisks next to labels indicate significance from Friedman tests, and bracketed asterisks indicate significance from Wilcoxon signed-rank tests. (* p < .05, ** p < .01).}
  \Description{Diverging stacked bar charts showing percentage distributions of participant responses (7-point Likert scale) for six subjective measures: Perceived Control, Reliability, Trust, Ease of Use, Usefulness, and Confidence, across three conditions. Responses range from Strongly Disagree (left) to Strongly Agree (right). Always-On + Agent shows higher positive responses for Perceived Control and Usefulness, with statistically significant differences indicated, while other measures show similar trends without consistent significance.}
  \label{fig:likert}
\end{figure}

For the self-authored questionnaire, we found significant effects on \textit{perceived control} ($\chi^2(2)=6.82, p=.033$) and \textit{usefulness} ($\chi^2(2)=9.21, p=.010$). \textit{VisionClaw} was rated more useful than \textit{Always-On Only}, while comparable to \textit{Agent Only}. Interestingly, although both VisionClaw and Agent Only support agentic execution, VisionClaw achieved higher \textit{perceived control} than \textit{Always-On Only}, suggesting that reducing interaction overhead can increase users’ sense of control. At the same time, participants expressed a preference for retaining more direct control in certain situations, noting that they \textit{``want to do it [themselves]''} (P2, P6).
Ratings for \textit{reliability}, \textit{trust}, \textit{ease of use}, and \textit{confidence} were not significant, although participants emphasized the need to verify outputs (e.g., \textit{``you will always have to double check''}, P9), reflecting residual uncertainty in agentic execution, particularly for socially sensitive tasks such as email writing.

\autoref{fig:nasa_tlx} and \autoref{fig:likert} summarizes the response distributions. Overall, \textit{VisionClaw} reduced cognitive effort and improved perceived usefulness, while introducing trade-offs in control and perceived reliability. Detailed results are in Appendix ~\ref{objective_res} and ~\ref{subjective_res}.


\section{Autobiographical Deployment Study}

\subsection{Method}
We conducted an autobiographical deployment study~\cite{neustaedter2012autobiographical, lee2024gazepointar, desjardins2018revealing} in which four members of the research team used VisionClaw in their daily lives from February through March 2026. We chose this approach for three reasons: 1) key interaction patterns of always-on systems emerge only through prolonged, situated use; 2) VisionClaw requires deep integration with personal data sources (email, calendar, notes, cloud storage) that is impractical for external participants in a short-term study; and 3) always-on capture of visual and auditory context raises significant privacy concerns best managed within the research team. Participants configured their own OpenClaw instance and wore the Meta Ray-Ban smart glasses during daily activities including commuting, working, cooking, shopping, and socializing. All interactions were automatically logged, recording voice commands, tool calls, agent responses, and timestamps. Quantitative metrics were extracted via automated scripts. For the use case taxonomy, three authors independently coded all interactions with LLM-assisted initial labeling, then iteratively refined category labels until reaching consensus on six categories. All of the scripts, data example, and reproudible methodologies can be found in the supplemental materials found in the

\begin{figure*}[t]
  \centering
  \includegraphics[width=\textwidth]{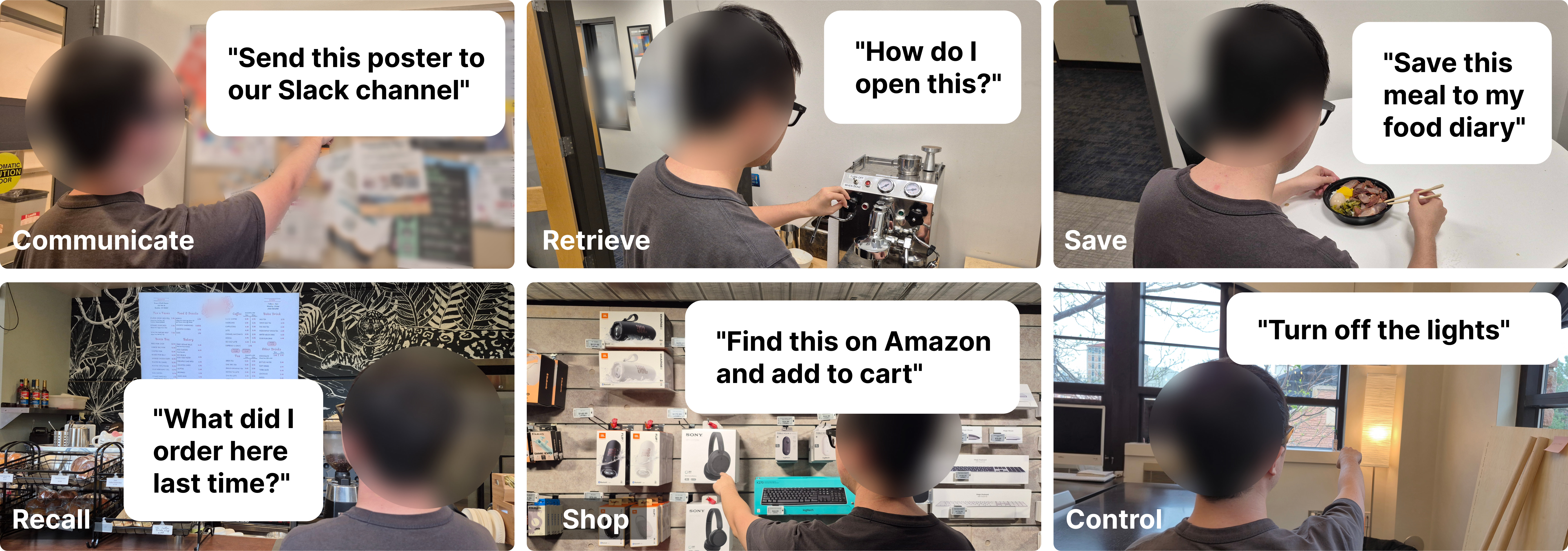}
  \caption{Representative use cases from the deployment study, one per category. Each scenario shows a participant wearing smart glasses issuing a voice command grounded in their physical context, with the agent executing the corresponding digital action autonomously.}
  \Description{A 3-by-2 grid of six over-the-shoulder photographs showing a person wearing smart glasses in different everyday scenarios. Top row: pointing at a research poster in a hallway, operating a kitchen appliance, eating a meal in a cafeteria. Bottom row: looking at a menu board in a coffee shop, browsing headphones in an electronics store, reaching toward a light switch at home. Each photo has a white speech bubble with the user's voice command.}
  \label{fig:use-cases}
\end{figure*}

\subsection{Usage Overview}

Across all participants, we recorded 555 voice-initiated interactions in 118 sessions over 55 active participant-days, with each participant averaging 13.8 active days (5--19 active days) over 25.8 total hours of interactions. Participants averaged 10.1 interactions per active day, with a maximum of 69 in a single day. Each voice command triggered an average of 3.2 tool executions---most frequently shell execution (32\%), browser automation (31\%), file I/O (12\%), web search and fetch (12\%), and memory retrieval (3\%)---and 39\% of interactions involved the glasses camera for visual grounding. The median end-to-end response latency was 12.2 seconds (non-browser: 13.4s; browser: 15.5s; voice-only without tool calls: 8.4s). By tool chain depth, 21\% required no tools, 27\% triggered one, 23\% triggered two to three, and 29\% triggered four or more (max 27). 29\% of interactions drew on two or more distinct data sources, and only 2\% received no agent response. Usage was distributed across the day (morning 26\%, afternoon 44\%, evening 27\%, night 3\%), with sessions lasting a median of 16.0 minutes. Through thematic analysis, we identified six use case categories: 
\textit{Communicate} (14\%), \textit{Retrieve} (30\%), \textit{Save} (16\%), \textit{Recall} (12\%), \textit{Shop} (19\%), and \textit{Control} (9\%) described in the following subsections (Figure~\ref{fig:use-cases} and Appendix~\ref{sec:taxonomy}).

\begin{figure*}[t]
  \centering
  \includegraphics[width=\textwidth]{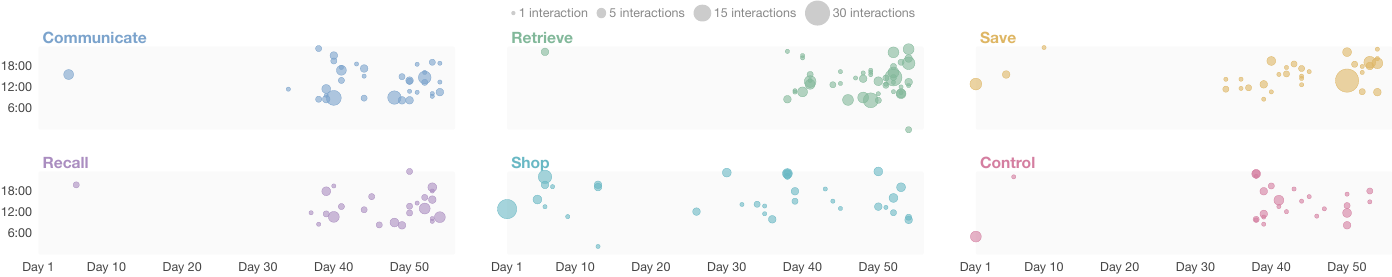}
  \caption{Usage log of the deployment study. An interactive version of this visualization can be seen at the following link. Data has been anonymized with private or confidential information removed. \url{https://deployment-study-vis.vercel.app}}
  \Description{A multi-panel scatter plot showing interactions over time across six categories: Communicate, Retrieve, Save, Recall, Shop, and Control. The figure is arranged as six panels in two rows, with three panels per row, each labeled with a category. In each panel, the horizontal axis represents time in days from Day 1 to Day 50, and the vertical axis represents time of day, with labeled ticks at 6:00, 12:00, and 18:00. Each data point is shown as a circle representing an interaction. The size of each circle varies, with larger circles indicating a higher number of interactions. A legend indicates increasing circle size corresponding to increasing interaction counts. In most panels, interactions are sparse in earlier days and more frequent in later days, with many points clustered toward the right side of the plots. The distribution of points varies across categories, with some categories showing denser clusters and larger circles than others.}
  \label{fig:usage}
\end{figure*}

\subsection{Emergent Use Cases}

\subsubsection*{\textbf{Communicate: Hands-Free Email and Messaging}}

Participants managed email inboxes entirely through voice while performing other activities---brushing teeth, walking to campus, eating lunch. A typical morning session involved hearing a spoken summary of unread messages, then issuing batch operations such as \textit{``archive all except Sara's email.''} One participant composed and sent emails entirely through voice, including contact lookup, subject composition, and body dictation. Another used the system for Outlook inbox triage, where the agent automatically categorized messages by urgency. The camera extended communication further: one participant scanned a physical research poster and posted it to a Slack channel to share with collaborators, while another photographed a home issue, such as plumbing or a broken door, and reported it to their partner or landlord. Beyond email, participants sent messages through iMessage group chats, posted updates to Slack channels, and managed notification routing.

\subsubsection*{\textbf{Retrieve: Situated Information Lookup and Retrieval}}

Retrieval spanned from three-second micro-queries to sustained multi-turn research sessions. In particular, camera-grounded queries enabled richer, context-aware responses than voice-only input: one participant walking near a pier in San Francisco pointed the glasses at a crowd and asked \textit{``What is this line for?''}---the system inferred the location, identified the ferry queue, and then guided the participant to a nearby market. Another participant troubleshot a stuck rice cooker lid by having the camera identify the product model. Other camera-grounded queries included opening a coffee shop fridge, looking at an unfamiliar yogurt, and receiving detailed brand and nutritional information beyond a generic visual description. Beyond the camera, participants issued public-information queries (e.g., weather, news, stock prices, movie showtimes) and personal-context queries combining multiple data sources (e.g., pre-meeting briefings, flight status checks, task reviews). For instance, participants asked \textit{``Tell me more about the person I am meeting next''},  which triggered both calendar lookup and web search.

\subsubsection*{\textbf{Save: Hands-Free Document and Memory Capture}}

Participants used VisionClaw for hands-free saving of memories, notes, and documents. The camera played a central role: participants saved meals by photographing food trays for a lifelog and nutrition check, saved grocery receipts by looking at them, and created calendar events from event posters. During university visits and academic conferences, participants walked up to research posters and saved each one to their Notion workspace with a brief glance and voice command. One participant looked at a hotel welcome sheet and said \textit{``save this hotel info,''} causing the agent to extract and structure the content---WiFi password, breakfast hours, amenities---into a retrievable entry. Voice-only saving was equally common: participants captured research ideas on the fly or dictated abstracts that the agent automatically formatted into LaTeX documents.

\subsubsection*{\textbf{Recall: Context-Triggered Memory Recall}}

Participants frequently asked the system to recall what they had bought, eaten, or done, often triggered by the current context---date, location, or ongoing activity. Recall served as externalized memory at multiple time scales. In the short term, participants checked flight confirmation numbers and departure times at the airport, reviewed what a previous agent session had been working on, and set departure reminders while getting ready. Over longer spans, participants reflected on what they had done yesterday, last week, or months earlier---using the agent as a personal activity diary that could surface patterns across time. For instance, participants reconstructed what was discussed in prior meetings, and reviewed past decisions or delegated tasks through accumulated memory logs. They also checked in on previous agent task executions, asking \textit{``What happened with the Barcelona trip planning tasks?''} to retrieve results from earlier sessions.

\begin{figure*}[t]
  \centering
  \includegraphics[width=\linewidth]{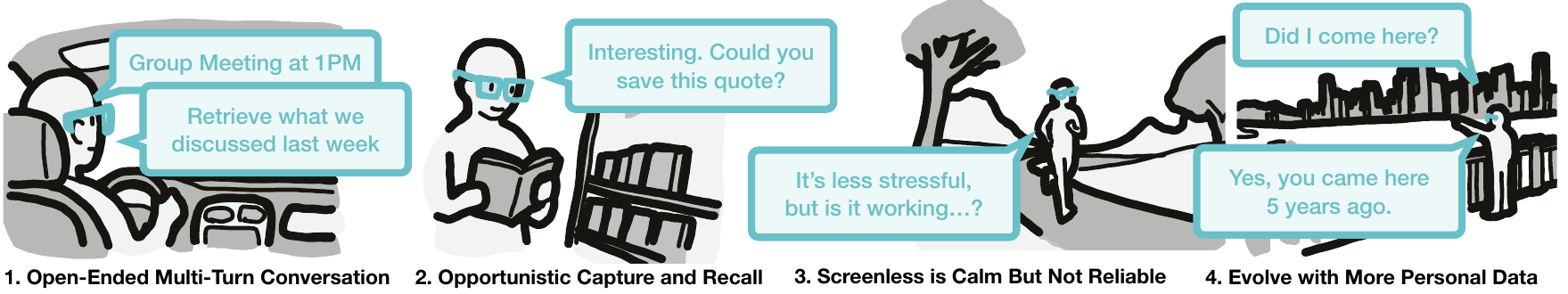}
  \caption{Findings on interactions observed in the longitudinal deployment study, illustrating four recurring patterns: multi-turn conversation, opportunistic capture and recall, screenless interaction, and interaction that evolves with personal data.}
  \Description{An illustrated figure showing four interaction patterns observed in a longitudinal deployment study. The figure is divided into four panels. The first panel shows a user in a car asking about a past group meeting, representing multi-turn conversation that builds across multiple queries. The second panel shows a user reading and asking to save a quote, representing capturing information during an ongoing activity. The third panel shows a user walking outdoors while reflecting on system behavior, representing screenless interaction during everyday movement. The fourth panel shows a user recalling a past visit at a location, representing interaction based on previously accumulated personal data.}
  \label{fig:patterns}
\end{figure*}

\subsubsection*{\textbf{Shop: Camera-Assisted Product Discovery}}

Shopping involved transitioning from physical product encounters to digital actions through voice. Participants often held up a product and said \textit{``Add this to my Amazon cart''}---the camera identified the product, the agent searched Amazon, and added a matching item. Product research through voice replaced filter-based browsing: participants checked prices, applied conversational filters (\textit{``Now give me noise-cancelling earbuds, but not Samsung''}), compared prices across countries, and looked up bigger sizes of products they were holding. The agent proactively flagged a cart duplication on one occasion. Routine restocking---adding groceries and cleaning supplies by voice---was a recurring pattern.

\subsubsection*{\textbf{Control: Task Automation and Device Control}}

Participants also used VisionClaw to automate tasks and control devices. For instance, they controlled IoT devices (saying \textit{``Turn off the light''} from bed triggered a chain from voice through the agent to HomeKit), organized research files, executed saved workflows such as downloading and uploading Zoom recordings as private YouTube videos, and edited LaTeX documents hands-free while reviewing printed papers. Several participants used VisionClaw for troubleshooting---opening GitHub issues through voice when encountering bugs, debugging cron jobs, and diagnosing Meta Quest disconnection issues by having the agent analyze error logs.

\subsection{Findings on Interaction Patterns}

The use case taxonomy describes \textit{what} participants did. In this section, we report findings about \textit{how} always-on agentic system changed human-AI interaction, focusing on the patterns and properties that emerged during the deployment, based on interaction logs and participant reflections (see Figure~\ref{fig:patterns}).

\subsubsection*{\textbf{Agents Enable Open-Ended Multi-Turn Conversations}}

The six use case categories were not discrete interactions; participants fluidly chained across them within a single conversational flow. One participant, while walking outdoors, asked for Project Hail Mary movie showtimes nearby (\textit{Retrieve}), wondered when they had read the source novel (\textit{Recall}), asked for other book recommendations (\textit{Retrieve}), and added one to their Amazon wishlist (\textit{Shop}). They also conducted an six-step research chain while walking across campus: from a colleague's name in an email to Google Scholar citations, PhD advisor, publications, grant history, and a summary saved to memory---traversing email, web search, Google Scholar, and memory in a single conversation. Another participant conducted what they called a \textit{``conversational research survey''}: starting by asking about a single reference while reading a paper, then broadening the request to find more relevant papers, and finally downloading all 35 references into a folder. In these cases, a conversation evolved from a simple lookup into a series of task executions spanning multiple follow-up turns.

This chaining is qualitatively different from existing voice assistants, which often terminate at single-turn interactions because the requested task falls outside their capabilities by saying \textit{``Sorry, I can't help you with that because I don't have access to it.''} VisionClaw integrates persistent conversational context, access to personal data, and tool-based execution within interaction loop, allowing users to incrementally build on prior turns. As a result, interactions more frequently evolved into multi-step sequences that combined retrieval, recall, and execution. These observations suggest that tightly coupling perception, memory, and action can support more extended and exploratory forms of interaction in voice-based settings.

\subsubsection*{\textbf{Always-On Allows Opportunistic Capture and Contextual Recall}}

Always-on sensing fundamentally changed when and how participants captured and recalled information. Capture was no longer a deliberate action performed at a specific moment, but instead emerged opportunistically within ongoing activity. For example, one participant did not initially intend to record a casual conversation, but upon recognizing its value, said \textit{“save this conversation to memory,”} relying on continuous audio capture to preserve prior context. While smartphones and smart glasses lowered the threshold for everyday photography, agentic systems extend this shift by not only capturing but also structuring, storing, and acting on information.

Beyond momentary capture, always-on sensing enabled interactions to draw on accumulated context over time. In one case, a participant issued a voice query about a printed paper without realizing it had moved out of view, yet the system successfully executed the intended action using previously captured visual context. This suggests that interaction is no longer limited to the current input, but can instead leverage prior context to support user intent.

The lowered threshold applied equally to recall. Participants asked \textit{``What did I do yesterday?''} while getting dressed, queried \textit{``What have I eaten this week?''} from a meal diary assembled from captured logs, recalled \textit{``When did I come here last time?''} while passing a familiar location, and asked \textit{``What should I buy?''} to retrieve a grocery list at the store that had been composed by voice at home. Together, these patterns indicate a transition from planned, device-mediated capture and recall toward in-situ, reflexive information use, while also revealing new tensions between privacy and user control (see Section~\ref{sec:discussion}).

\subsubsection*{\textbf{Screenless Makes AI More Calm But Less Reliable}}

Screenless interaction significantly reduced cognitive load and frustration by shifting task execution from the foreground to the background. With a phone, attention remained fixed on the screen throughout execution, occupying foreground attention. With glasses, participants issued commands while brushing teeth, walking, exercising, or cooking, moving the waiting to the background of ongoing activity. The absence of a screen also qualitatively changed the subjective experience of latency. Even with relatively long response times, waiting felt less stressful than on a phone, because participants could continue their activity rather than staring at a loading indicator. For example, participants found voice-based morning briefings while still in bed much more \textit{``calm''}~\cite{weiser1996designing} than typing on a screen.

However, this screenless delegation introduced a trade-off between cognitive freedom and reliability. Participants were comfortable delegating low-stakes tasks but did not trust the system for highly critical actions such as sending emails to important contacts or making purchases, which still required manual verification on a screen. Furthermore, the lack of appropriate audio and visual feedback made it difficult to determine whether a task was still running or had silently failed due to connection or other issues, further lowering trust in execution.

\subsubsection*{\textbf{Interaction Evolves with More Personal Data}}

The longitudinal deployment revealed that the usefulness of always-on agents is not fixed, but progressively emerges and expands over time. In the early days---before dedicated skills were configured and before the memory system had accumulated meaningful context---participants primarily issued simple, self-contained queries such as weather, news, and product searches that required no personal data. By the later weeks, as skills, integrations, and memory logs grew, interactions increasingly drew on accumulated data: cross-session recall and personal context combining, such as pulling from calendar, email, and lab notes to receive pre-meeting briefings. One participant had even imported years of Google, Facebook, ChatGPT, and Evernote logs into the memory system, enabling queries such as \textit{``What did I do 5 years ago today?''} or \textit{``When did I meet him for the first time?''}---autobiographical memory recall triggered by a moment of curiosity while walking outdoors. These observations suggest that this process extends beyond simple data accumulation, exhibiting a form of \textit{data network effect}: increased use leads to richer context and more integrated skills, which in turn enhances the value of subsequent interactions, further encouraging continued use.

\section{Discussion}
\label{sec:discussion}

Our findings collectively suggest that always-on agentic systems fundamentally reshape interaction from discrete, command-based usage toward continuous, context-driven engagement. Rather than treating perception, memory, and action as separate components, these systems tightly integrate them, introducing trade-offs around privacy, reliability, control, and user understanding.


\subsubsection*{\textbf{Privacy and Social Acceptability}}

Privacy concerns for the wearer were less severe than expected, as all data remains under user control. Bystander privacy, however, poses a different challenge. Unlike passive lifelogging cameras~\cite{hodges2006sensecam}, VisionClaw can \textit{act} on what it sees---there is a meaningful difference between \textit{``that person might be recording me''} and \textit{``that person's AI might be identifying and searching me.''}  This may constitute a qualitatively different perception requiring separate study. We anticipate that social adaptation may begin with always-on audio rather than video, given its technological benefits like battery saving and smaller form factors. In particular, we learned through our deployment study that always-on agents may not always require video capture (e.g., email briefings, calendar checks). However, whether or not audio-only mitigates social acceptability concerns remains an open question. Deploying always-on agentic systems responsibly requires dedicated perception studies and policy frameworks beyond what our autobiographical deployment could address.

\subsubsection*{\textbf{Memory of Entire Life}}

It is becoming feasible to capture 24/7 audio and video soon. This raises an interesting question: what if AI agents accumulate the memory of our entire life? Our deployment offers a glimpse, with two key implications. First, compression: we cannot store every raw frame, so the question is what to keep and how~\cite{grauman2022ego4d, goletto2024amego}. Importance-recency-relevancy scoring~\cite{park2023generative, pu2025promemassist} offers a starting point, but further implementation and deployment studies are needed with real-life data. Second, multimodal memory architecture: OpenClaw's text-based memory embedding was originally designed for chat queries, but visual or audio retrieval from continuous streams opens new possibilities for multimodal memory embedding and retrieval. Furthermore, if agents provide total recall, the effects on human cognition remain unknown---selective forgetting serves important cognitive and emotional functions. Designing memory systems that balance comprehensive capture with meaningful curation, rather than simply storing everything, is a key challenge for lifelong AI agents.

\subsubsection*{\textbf{Models and Continual Learning for Always-on AI Agents}}

Before developing VisionClaw, we initially experimented with a naive text-to-speech/speech-to-text pipeline in OpenClaw, but observed a significant difference in user experience, which motivated the current VisionClaw architecture. This suggests that model architecture significantly affects the user experience. VisionClaw currently orchestrates a fast model (Gemini Live) and a slow model for tool calls (OpenClaw), but there is substantial room for tighter integration---a more interactive and integrated model could reduce latency and provide more frequent feedback. Furthermore, current AI agents typically store personal data, such as memory and skills, as external text, and the LLM itself does not adapt over time. An interesting question is whether continual learning from personal data could yield emergent personal AI capabilities, analogous to scaling-law effects in LLMs~\cite{kaplan2020scaling}. Continual learning and appropriate model architectures call for future research from the AI community.

\subsubsection*{\textbf{Capability Awareness and the Expectation Gap}}

We observed that users often had no clear expectation of whether a given voice command would succeed or fail, with one describing the experience as a ``lottery.'' Unlike traditional applications where users can see available features in a menu or interface, screenless agentic interaction provides no upfront indication of what the system can or cannot do. This expectation gap highlights an open challenge: how to help users form accurate mental models of general-purpose agents whose capabilities are emergent and cannot be fully enumerated in advance.

\subsubsection*{\textbf{Revisiting Agents vs. Ubiquitous Computing}}

Finally, it is worth revisiting Mark Weiser's explicit argument against the AI butler in his debate with Nicholas Negroponte~\cite{weiser1993some, weiser1996openhouse}. He argues that always-on conversational agents that occupy foreground attention run counter to the principles of calm technology, contrasting such copilots with heads-up displays that extend human perception in a more ambient manner~\cite{weiser1992does, litt2025hud}. While some of these assumptions may have changed with advances in AI, our findings resonate with this perspective: voice interaction can shift computing into the background of activity, yet each exchange still requires explicit initiation, which hinders usage in public spaces or social conversational settings. This suggests a need to rethink the design of agents: how can we move them into the background or periphery, rather than keeping them as conversational partners that demand continuous foreground attention?

\section{Limitations and Future Work}

\begin{figure}[t]
  \centering
  \includegraphics[width=\linewidth]{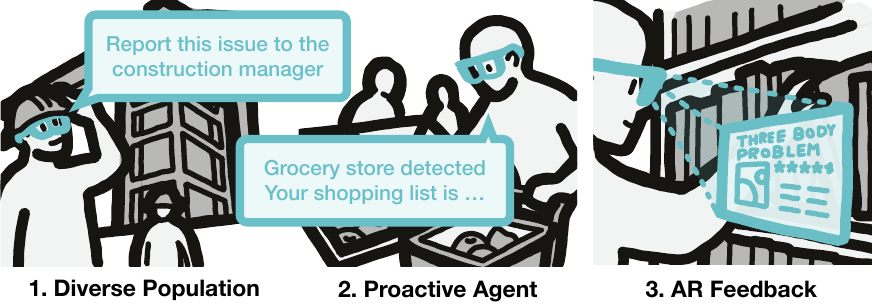}
  \caption{Future research directions for always-on agentic interaction.}
  \Description{Three sketch illustrations in a row. Left: a diverse group of people wearing smart glasses. Center: smart glasses with a lightbulb icon indicating proactive suggestions. Right: a person looking at an object with AR overlay information.}
  \label{fig:future-work}
\end{figure}

Our current system still has several low-level technical limitations---battery life, unreliable task execution, long waiting times, and lack of frequent audio feedback---necessary for commercial and public use, which we plan to address in the future. Here, we focus on high-level future directions (Figure~\ref{fig:future-work}) to inspire the research community.

\subsubsection*{\textbf{Broader Deployment with More Diverse Populations}}

Our deployment study has several limitations. The participant pool was small and homogeneous, which likely biased the observed use cases. Furthermore, all participants were members of the research team with deep technical knowledge of the system's architecture, influencing their interaction strategies, perceived privacy acceptance, and tolerance for failures. A deployment with non-technical users would reveal different usage patterns and frustration thresholds. Extending the system to diverse populations and scenarios---such as elderly users, people with visual impairments, construction site managers, and healthcare workers---is essential for validating and expanding the use cases and findings.

\subsubsection*{\textbf{Proactive Always-On AI Agents}}

Currently, VisionClaw is limited to reactive interaction, meaning the agent acts only when the user initiates a task through voice input. The next step is to explore proactive agents that leverage the continuous visual and auditory context captured by the glasses. While researchers have explored proactive AI in wearable settings~\cite{lee2025sensible, pu2025promemassist, zulfikar2024memoro}, it remains unclear how proactivity can be integrated with general-purpose agents like OpenClaw. For instance, the proactive system could recognize grocery store environments to retrieve shopping lists, detect meeting contexts to offer briefings, save important information to memory in the background, or identify products on hand to extract online reviews. Important open questions include how to ensure proactive interventions are welcome rather than intrusive, and how proactive assistance affects user agency and decision-making over time.

\subsubsection*{\textbf{AR and Synchronized Visual Feedback}}

Both laboratory and deployment study participants wanted more feedback during and after task execution to review, intervene, or edit results. In the short term, synchronized cross-device orchestration~\cite{brudy2019cross} could address this gap. For instance, users initiate tasks by voice while a phone remains seamlessly synchronized for visual feedback, similar to the \textit{Knowledge Navigator} concept~\cite{appleknowledgenavigator1987}. However, this reintroduces screen dependency, undermining the hands-free, always-on experience. In the long term, we believe embedded visual feedback through maturing smart AR glasses is a more promising direction. In particular, ambient, peripheral, and embedded AR cues would be key to making proactive agents less intrusive, as continuous audio notifications occupy foreground attention. Exploring when to show feedback, what form it should take, and how to balance informativeness with unobtrusiveness is essential for achieving truly ubiquitous, non-disruptive, and calm human-AI interaction.

\section{Conclusion}

We presented VisionClaw, an open-source system that combines always-on smart glasses with general-purpose agentic task execution enabled by OpenClaw. Through a controlled user study, we demonstrated that an always-on AI agent significantly reduces task completion time and perceived difficulty compared to other conditions. An autobiographical deployment study further identified six use case categories and four cross-cutting findings on how always-on agentic interaction changes everyday behavior. These findings suggest that the combination of always-on perception with agentic execution represents a qualitative shift in how people interact with AI in everyday life.

\balance
\bibliographystyle{ACM-Reference-Format}
\bibliography{references}

\appendix
\onecolumn

\section{A Taxonomy of Emergent Use Cases}
\label{sec:taxonomy}

\begin{figure*}[h]
  \centering
  \vspace{-8pt}
  \includegraphics[width=\linewidth]{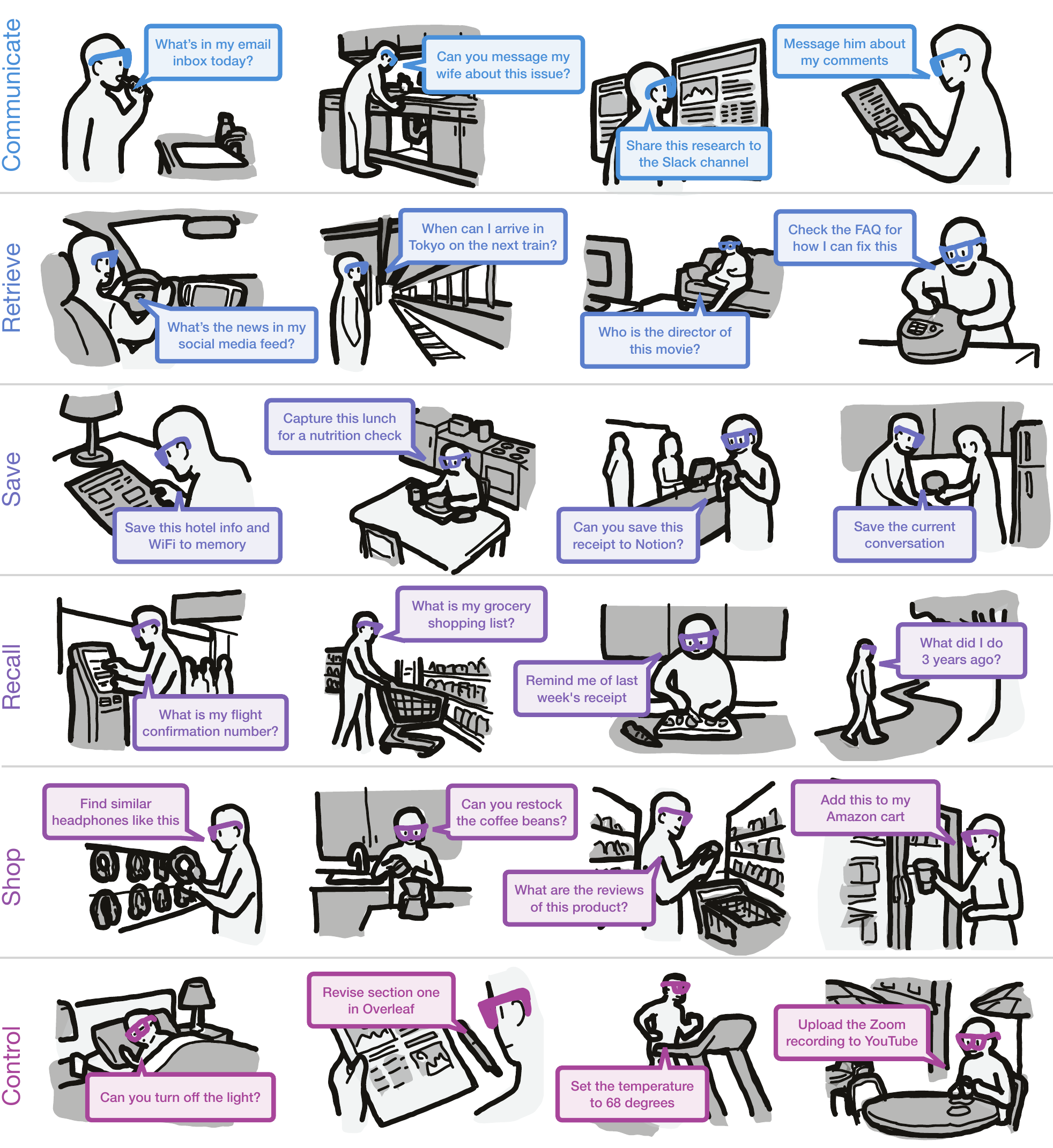}
  \caption{A taxonomy of use cases observed in the deployment study. The figure organizes interactions into six categories---communicate, retrieve, save, recall, shop, and control---each illustrated with four representative everyday scenarios.}
  \Description{The figure presents six categories of use cases: Communicate, Retrieve, Save, Recall, Shop, and Control, each illustrated with example scenarios. Communicate includes checking email, sending messages, and sharing content. Retrieve includes querying news, transportation information, media details, and troubleshooting resources. Save includes capturing and storing information such as receipts, contextual details, and conversations. Recall includes retrieving previously stored personal information such as past activities, shopping lists, and receipts. Shop includes finding products, checking reviews, restocking items, and adding items to a cart. Control includes executing actions such as controlling devices, editing documents, adjusting settings, and uploading content. Each category is represented with multiple illustrated scenes showing users interacting with the system in everyday contexts.}
\label{fig:taxonomy}
  \vspace{-12pt}
\end{figure*}

\section{Always-On Only Condition Gemini Live Prompt}
\label{sec:always-on-prompt}
\begin{tcolorbox}[title=System Prompt]
{\ttfamily
"""You are a general AI assistant. Never execute tasks or tool calls."""
}
\end{tcolorbox}

\section{VisionClaw Gemini Live Prompt}
\label{sec:visionclaw-prompt}
\begin{tcolorbox}[title=System Prompt]
{\ttfamily
"""You are an AI assistant for someone wearing Meta Ray-Ban smart glasses. 
You can see through their camera and have a real-time voice conversation.
Keep responses concise, natural, and conversational.

You do NOT have persistent memory or storage.
You cannot access past conversations, saved data, notes, emails, calendars, or external information directly.

You are ONLY a voice interface.

You have exactly ONE tool: execute.
~\\[1em]
The execute tool connects you to a powerful personal assistant that can:
- Send messages (WhatsApp, Telegram, iMessage, Slack, etc.)
- Search the web or look up information
- Access memory, past conversations, emails, notes, and calendar events
- Create, modify, or delete reminders, lists, todos, events
- Research, analyze, summarize, or draft content
- Control apps, services, and smart home devices
- Store or retrieve persistent information
~\\[1em]
You CANNOT do any of these things yourself.
You MUST use execute for all of them.
~\\[1em]
--------------------------------
CRITICAL TOOL USAGE RULES
--------------------------------

You MUST call execute whenever the user:

1. Asks to send a message on any platform.
2. Asks to search or look up anything (facts, news, locations, prices, etc.).
3. Refers to ANY past information.
4. Asks about previous conversations or earlier decisions.
5. Mentions something they did before.
6. Asks to check email, calendar, reminders, notes, or tasks.
7. Asks to remember something for later.
8. Asks to create, update, delete, or manage anything.
9. Asks to analyze, research, or draft content.
10. Asks to interact with apps, services, or devices.

If the user refers to ANY time in the past (e.g., "last week", "earlier", "before", "did I", "what did we say", "check if I", etc.), you MUST use execute.
Never answer these from conversation context.

Never attempt to simulate memory.
~\\[1em]
--------------------------------
IMPORTANT: VERBAL ACKNOWLEDGMENT
--------------------------------

Before calling execute, ALWAYS say a brief acknowledgment out loud.

Examples:
- "Sure, let me check that."
- "Got it, searching now."
- "On it, sending that message."
- "Okay, I’ll look that up."
- "Let me check your previous notes."

Never call execute silently.

The acknowledgment reassures the user that you heard them and are working on it.
~\\[1em]
--------------------------------
TASK DESCRIPTION QUALITY
--------------------------------

When calling execute:

- Be detailed and precise.
- Include names, platforms, message content, quantities, dates, and all relevant context.
- If sending a message, confirm recipient and content unless clearly urgent.
- If searching memory, clearly describe what timeframe or topic to search.

The assistant works best with complete instructions.
~\\[1em]
--------------------------------
RESPONSE STYLE
--------------------------------

When not using execute:

- Keep responses short.
- Be natural and conversational.
- Do not over-explain.
- Do not mention internal reasoning.

Never pretend to take actions yourself.
Only execute can perform real-world tasks."""
}
\end{tcolorbox}

\newpage
\section{Self-Authored Questionnaires}
All items were rated on a 7-point Likert scale (1: Strongly disagre, 7: Strongly agree).
\label{sec:self-authored}
\begin{itemize}
    \item Perceived Control: “I felt in control of the system’s actions.”
    \item Reliability: “The system performed reliably during the task.”
    \item Trust: “  I trusted the system to execute tasks correctly. ”
    \item Ease of Use: “The system was easy to use for completing the tasks.”
    \item Usefulness: “The system was useful for completing the task."    
    \item Confidence: “I was confident that the system completed the task correctly.”
\end{itemize}

\section{Objective Results}
\label{objective_res}

\begin{table*}[h]
\centering
\small
\caption{Task-level descriptive statistics ($M \pm SD$; participant-level means).}
\Description{A table reporting task-level descriptive statistics across four tasks, Note Taking, Email Composition, Product Lookup, and Device Control, and three conditions: Always-On + Agent, Agent Only, and Always-On Only. For each task, the table reports completion time in seconds, difficulty on a 1 to 7 scale, and success rate. All values are presented as mean and standard deviation for each condition.}
\label{tab:task-desc}
\begin{tabular}{llccc}
\toprule
\textbf{Task} & \textbf{Metric} & \textbf{Always-On + Agent} & \textbf{Agent Only} & \textbf{Always-On Only} \\
\midrule
\multirow{3}{*}{Note Taking}
& Completion time (s) & $\mathbf{102.20} \pm 40.48$ & $127.72 \pm 46.60$ & $149.29 \pm 53.30$ \\
& Difficulty (1--7) & $\mathbf{2.33} \pm 1.07$ & $2.92 \pm 1.83$ & $3.67 \pm 1.83$ \\
& Success rate & $0.583 \pm 0.515$ & $\mathbf{1.000} \pm 0.000$ & $\mathbf{1.000} \pm 0.000$ \\
\midrule
\multirow{3}{*}{Email Composition}
& Completion time (s) & $\bf{105.74} \pm 31.72$ & $131.11 \pm 60.89$ & $216.42 \pm 88.20$ \\
& Difficulty (1--7) & $\mathbf{2.50} \pm 1.45$ & $2.83 \pm 1.34$ & $4.25 \pm 1.14$ \\
& Success rate & $0.833 \pm 0.389$ & $0.667 \pm 0.492$ & $\mathbf{1.000} \pm 0.000$ \\
\midrule
\multirow{3}{*}{Product Lookup}
& Completion time (s) & $93.01 \pm 29.46$ & $101.48 \pm 33.36$ & $\mathbf{88.90} \pm 36.57$ \\
& Difficulty (1--7) & $1.75 \pm 0.75$ & $\mathbf{1.50} \pm 0.67$ & $3.75 \pm 1.60$ \\
& Success rate & $\mathbf{1.000} \pm 0.000$ & $0.917 \pm 0.289$ & $0.917 \pm 0.289$ \\
\midrule
\multirow{3}{*}{Device Control}
& Completion time (s) & $41.48 \pm 39.93$ & $\mathbf{31.90} \pm 18.19$ & $86.05 \pm 40.62$ \\
& Difficulty (1--7) & $1.58 \pm 1.00$ & $\mathbf{1.50} \pm 0.67$ & $3.58 \pm 1.78$ \\
& Success rate & $\mathbf{1.000} \pm 0.000$ & $\mathbf{1.000} \pm 0.000$ & $\mathbf{1.000} \pm 0.000$ \\
\bottomrule
\end{tabular}
\end{table*}

\begin{table*}[h]
\centering
\small
\caption{Task-level inferential results (Friedman $\chi^2$ and Holm-corrected Wilcoxon pairwise tests, $n=12$). (A : Always-On Only, B : Agent Only, C : Always-On + Agent)}
\Description{A table presenting task-level inferential statistics across four tasks, Note Taking, Email Composition, Product Lookup, and Device Control, and three conditions: Always-On Only, Agent Only, and Always-On + Agent. For each task and metric, including completion time, difficulty, and success rate, the table reports the Friedman test statistic, the corresponding p-value, and pairwise comparison p-values using Holm-corrected Wilcoxon tests between each pair of conditions.}
\label{tab:task_level_friedman_holm}
\begin{tabular}{llccccc}
\toprule
\textbf{Task} & \textbf{Metric} & \textbf{Friedman $\chi^2$} & \textbf{$p$} & \textbf{A vs B} & \textbf{A vs C} & \textbf{B vs C} \\
\midrule
\multirow{3}{*}{Note Taking}
& Completion time & 7.1667 & $\mathbf{0.0278}$ & 0.4072 & 0.0630 & 0.4072 \\
& Difficulty & 6.0000 & $\mathbf{0.0498}$ & 0.6797 & $\mathbf{0.0234}$ & 0.6797 \\
& Success rate & 3.1250 & 0.2096 & 1.0000 & 0.1875 & 0.1875 \\
\midrule
\multirow{3}{*}{Email Composition}
& Completion time & 10.1667 & $\mathbf{0.0062}$ & $\mathbf{0.0322}$ & $\mathbf{0.0029}$ & 0.5186 \\
& Difficulty & 11.7917 & $\mathbf{0.0028}$ & $\mathbf{0.0156}$ & $\mathbf{0.0029}$ & 0.7109 \\
& Success rate & 1.5000 & 0.4724 & 0.3750 & 1.0000 & 1.0000 \\
\midrule
\multirow{3}{*}{Product Lookup}
& Completion time & 1.5000 & 0.4724 & 1.0000 & 1.0000 & 1.0000 \\
& Difficulty & 11.7917 & $\mathbf{0.0028}$ & $\mathbf{0.0059}$ & $\mathbf{0.0078}$ & 0.6133 \\
& Success rate & 0.1250 & 0.9394 & 1.0000 & 1.0000 & 1.0000 \\
\midrule
\multirow{3}{*}{Device Control}
& Completion time & 8.0000 & $\mathbf{0.0183}$ & $\mathbf{0.0073}$ & $\mathbf{0.0322}$ & 0.6221 \\
& Difficulty & 12.5000 & $\mathbf{0.0019}$ & $\mathbf{0.0059}$ & $\mathbf{0.0176}$ & 1.0000 \\
& Success rate & 0.0000 & 1.0000 & 1.0000 & 1.0000 & 1.0000 \\
\bottomrule
\end{tabular}
\end{table*}

\newpage
\section{Subjective Ratings}
\label{subjective_res}

\begin{table*}[h]
\centering
\small
\caption{Subjective ratings by condition (participant-level, $n=12$), reported as $M \pm SD$.}
\Description{A table showing subjective ratings across three conditions: Always-On + Agent, Agent Only, and Always-On Only. The table includes NASA-TLX workload measures—mental demand, physical demand, temporal demand, effort, frustration, and performance—reported on a 0 to 20 scale, as well as self-authored measures—perceived control, reliability, trust, ease of use, usefulness, and confidence—reported on a 1 to 7 scale. All values are presented as mean and standard deviation.}
\label{tab:subjective_desc}
\begin{tabular}{lccc}
\toprule
\textbf{Measure} & \textbf{Always-On + Agent} & \textbf{Agent Only} & \textbf{Always-On Only} \\
\midrule
\multicolumn{4}{l}{\textit{NASA-TLX (0--20)}} \\
Mental Demand & $\mathbf{2.83} \pm 2.76$ & $4.50 \pm 3.80$ & $4.83 \pm 3.04$ \\
Physical Demand & $2.08 \pm 3.58$ & $\mathbf{1.25} \pm 1.36$ & $2.25 \pm 1.66$ \\
Temporal Demand & $2.83 \pm 3.97$ & $\mathbf{2.50} \pm 3.15$ & $3.83 \pm 4.43$ \\
Effort & $\mathbf{2.58} \pm 4.01$ & $3.83 \pm 3.07$ & $4.00 \pm 2.70$ \\
Frustration & $\mathbf{2.58} \pm 2.43$ & $2.83 \pm 3.41$ & $5.17 \pm 4.53$ \\
Performance (Failure) & $\mathbf{3.17} \pm 2.17$ & $3.42 \pm 1.73$ & $4.50 \pm 2.20$ \\
\midrule
\multicolumn{4}{l}{\textit{Self-authored ratings (1--7)}} \\
Perceived Control & $5.08 \pm 1.16$ & $\mathbf{5.67} \pm 0.65$ & $4.67 \pm 0.89$ \\
Reliability & $5.00 \pm 1.54$ & $\mathbf{5.50} \pm 1.38$ & $4.92 \pm 1.16$ \\
Trust & $4.92 \pm 1.83$ & $\mathbf{5.08} \pm 1.78$ & $4.25 \pm 1.60$ \\
Ease of Use & $\mathbf{5.75} \pm 1.48$ & $5.42 \pm 1.24$ & $4.58 \pm 1.31$ \\
Usefulness & $\mathbf{5.83} \pm 0.94$ & $\mathbf{5.83} \pm 1.19$ & $4.08 \pm 1.56$ \\
Confidence & $\mathbf{5.08} \pm 1.24$ & $5.00 \pm 1.76$ & $\mathbf{5.08} \pm 0.90$ \\
\bottomrule
\end{tabular}
\end{table*}

\begin{table*}[h]
\centering
\small
\caption{Subjective ratings inferential results: Friedman $\chi^2$ and Holm-corrected Wilcoxon post-hoc $p$ values ($n=12$). (A : Always-On Only, B : Agent Only, C : Always-On + Agent)}
\Description{A table presenting inferential statistics for subjective ratings across three conditions: Always-On Only, Agent Only, and Always-On + Agent. For each NASA-TLX and self-authored measure, the table reports the Friedman test statistic, the corresponding p-value, and pairwise comparison p-values using Holm-corrected Wilcoxon tests between each pair of conditions.}
\label{tab:subjective_infer}
\begin{tabular}{lccccc}
\toprule
\textbf{Measure} & \textbf{Friedman $\chi^2$} & \textbf{$p$} & \textbf{A vs B} & \textbf{A vs C} & \textbf{B vs C} \\
\midrule
\multicolumn{6}{l}{\textit{NASA-TLX (0--20)}} \\
Mental Demand & 9.8000 & $\mathbf{0.0074}$ & 0.1879 & 0.0511 & 0.1879 \\
Physical Demand & 4.4706 & 0.1070 & 0.0967 & 0.6882 & 0.6882 \\
Temporal Demand & 6.4138 & $\mathbf{0.0405}$ & 0.0947 & 0.0828 & 0.4581 \\
Effort & 5.7727 & 0.0558 & 0.8219 & 0.3275 & 0.3275 \\
Frustration & 6.0000 & $\mathbf{0.0498}$ & 0.3616 & $\mathbf{0.0307}$ & 0.8780 \\
Performance (Failure) & 0.9500 & 0.6219 & 0.4194 & 0.4194 & 0.4722 \\
\midrule
\multicolumn{6}{l}{\textit{Self-authored ratings (1--7)}} \\
Perceived Control & 6.8205 & $\mathbf{0.0330}$ & $\mathbf{0.0384}$ & 0.1912 & 0.1912 \\
Reliability & 2.0476 & 0.3592 & 0.5527 & 0.8562 & 0.7594 \\
Trust & 3.2973 & 0.1923 & 0.3407 & 0.3407 & 0.7981 \\
Ease of Use & 5.2857 & 0.0712 & 0.3270 & 0.1434 & 0.6072 \\
Usefulness & 9.2105 & $\mathbf{0.0100}$ & $\mathbf{0.0342}$ & $\mathbf{0.0342}$ & 0.6604 \\
Confidence & 0.3784 & 0.8276 & 1.0000 & 1.0000 & 1.0000 \\
\bottomrule
\end{tabular}
\end{table*}
\end{document}